# MASS BOOM VERSUS BIG BANG : THE ROLE OF PLANCK´S CONSTANT

by

Antonio Alfonso-Faus

E.U.I.T. Aeronáutica

Plaza Cardenal Cisneros s/n

28040 Madrid, SPAIN

e-mail: aalfonso@euita.upm.es

The Big Bang frame of work for cosmology is a theoretical construct based upon one possible interpretation of the Hubble observation of the red shift from distant galaxiwa. Almost all of the scientific experimental evidence has been interpreted under the Big-Bang hypothesis. And the results have been very promising though with a few important drawbacks. The main point in our present work is to prove that the expansion of the Universe, the essence of the Big-Bang, is based upon an apparent interpretation of the distance-red shift relation found by Hubble. On the contrary, an absolute interpretation as we give here results in a constant size Universe, together with a shrinking quantum world. This is the role of Planck´s constant: the red shift-distance relation can be interpreted as evidence of an expanding Universe or, alternatively, as evidence as evidence of a shrinking quantum world due to the decrease of Planck´s constant with cosmological time.

When considering possible time variations of fundamental physical constants one has to keep firm well established principles. Following this approach we keep firm the Action Principle, General Relativity (the Equivalence Principle), and Mach's Principle. Also we introduce a new principle under the name of "TOTAL INTERACTION" and reconsider Weinberg's relation with a new approach. We find that all masses increase linearly with cosmological time (THE MASS BOOM). The speed of light turns out to be decreasing also with time. An "absolute" cosmological model arises, similar to the one Einstein proposed, static, closed and finite, with the cosmological constant included. The "relative" model, the Universe as seen from the Lab observers, is an expanding one with a quadratic law in time for the cosmological scale factor, a model that fits all recent observations of dark mass and dark energy contents in the Universe, as well as the Supernova Type Ia data.

1.- INTRODUCTION

The beginning of scientific cosmology can be placed with the advent of the cosmological equations of Einstein, as derived from his theory of general relativity. A Universe with masses, and therefore gravitation, would be expected to contract, and that was the case initially predicted. To avoid this, and have a static solution for the Universe as a whole, Einstein included in his equations a cosmological constant, the lambda term, that resulted in a push to balance gravitation. Then he obtained a static model, with curved space giving a closed as well as finite Universe. Later Hubble found the redshift from distant galaxies to increase with distance. One of the possible explanations was to consider that the Universe was expanding as seen from any observer. Of course, going backwards in time, with this interpretation the Universe would be seen as contracting, therefore would be initially "born" from a relatively small size and much hotter and denser than now. An expanding Universe, found as a solution to the Einstein´s equations, was proposed by Friedman and therefore the scientific community had a theoretical frame to explain one interpretation of Hubble´s discovery. The concept of an expanding Universe from an initial Big Bang spread rapidly and has been considered the best model, and therefore the best frame, to interpret all current observations.

Nevertheless this model has many problems and contains many paradoxes that have been in part solved with additional theoretical inclusions (e.g. inflation).

In the present work we introduce the concept of a Mass Boom, already present in the literature[1] two years ago at the IV International Congress in Hyderabad, India, but now we present here important modifications and refinements. Keeping first principles firm we prove that all masses grow (increase) linearly with cosmological time. Conservation of momentum implies then that the speed of light decreases linearly with time. It is seen that Mach´s principle, and its equivalent the principle of equivalence, give a unique solution for the Universe that excludes expansion. We then reinterpret the redshift found by Hubble, and prove that at the laboratory system it is proportional to Planck´s "constant", a result that comes from the comparison of frequencies. The new interpretation that we present here depends exclusively on Planck´s "constant", that we find decreasing with cosmological time. The constancy of Planck´s units of time and mass completely defines the time variation of the Planck´s constant. The new model we propose for the Universe is a static, closed and finite one, as Einstein initially proposed. On the other hand we find a quantum world contracting with cosmological time, in accordance with the decrease of Planck´s constant, as interpreted from the redshift. It is clear that if we take the quantum world as the reference, then the Universe would be apparently expanding. It turns out that this apparent expansion is an accelerated one, something already observed from the supernova type Ia measurements[2]. The apparent expansion we find is $a(t) \propto t^2$, for the cosmological scale factor $a(t)$.

We then solve the cosmological equations, and find the corresponding numerical values for the dimensionless matter parameter $\Omega_m = 1/3$ and the lambda parameter $\Omega_\Lambda = 2/3$, which are very close to the values observed in many experiments. It turns out that the apparent curvature term becomes rapidly negligible with age, so that practically we live in a flat Universe, as seen from our Lab, also a well known current observation. The entropy of the Universe is found to be equivalent in some way to the cosmological time, as well as to the total matter of the Universe. Then, we can talk of a Mass Boom as well as an Entropy Boom, equivalent to cosmological time.

The natural units that emerge from this model, the "true constants of nature" are Planck´s mass and time, and of course the present size of the Universe $ct \approx 10^{28}$ cms, which is the Planck´s length at the first "tic" of the Universe (at the Planck´s time).

Finally, the Pioneer 10/11 anomalous acceleration[3] recently observed is explained by our theory presented here. We take this as an experimental evidence of the Mass Boom effect.

2.- THE MASS BOOM, PREDICTED BY FIRST PRINCIPLES.

There have been doubts whether general relativity included Mach´s principle or not. Certainly it includes the equivalence principle, and now we will present an interpretation of both principles that proves them to be equivalent. One interpretation of Mach´s principle considers the mass (energy) of a particle m as due to its gravitational potential energy with respect to the mass $M_u$ of the rest of the Universe

$$\frac{GM_u m}{ct} \approx mc^2 \qquad (1)$$

General relativity is based upon the equivalence principle. One way to express it in mathematical terms is to preserve, under any sort of time-variations, the ratio of the square of any speed due to gravitation, $v^2 = GM/r$, to the square of the speed of light $c^2$, i.e.

$$\frac{v^2}{c^2} = \frac{GM}{c^2 r} = const. \qquad (2)$$

The constancy of this ratio ensures the preservation of the principle of relativity under cosmological time variations. If we substitute for the size r the size of the seeable Universe, ct, and for M the mass of the Universe $M_u$ , one gets

$$\frac{GM_u}{c^3 t} = const. \qquad (3)$$

We see that the expressions (1) and (3) are equivalent. In the next section on the action principle we prove that G and $c^3$ have to be proportional to preserve the standard form of the field equations of general relativity. The result is that the mass of the Universe has to be proportional to the cosmological time (the Mass Boom) :

$$M_u = \text{constant} \cdot t \qquad (4)$$

We present now what we call the total interaction principle. It is a mathematical expression that follows the requirement that all the gravitational interactions in the Universe must have a mean free path, under a Newtonian point of view, of the order of the size of the Universe. Then,

$$ct \approx \frac{1}{n\sigma_g} \qquad (5)$$

where n is the number density of particles in the Universe and $\sigma_g$ their gravitational cross section as defined elsewhere[4] and given by

$$\sigma_g = 4\pi \frac{Gm}{c^2} \cdot ct \qquad (6)$$

Substituting the above into (5) one has

$$ct \approx \frac{(ct)^3}{\frac{G}{c^2} M_u ct} \qquad (7)$$

i.e.
$$ct \approx \frac{GM_u}{c^2} \qquad (8)$$

which is the same as (3), the equivalence principle, and the same as (1), the Mach's principle.

Finally, by using the mass of the quantum of gravity $m_g$ defined elsewhere[5] as

$$m_g = \frac{\hbar}{c^2 t} \qquad (9)$$

and calculating the mass rate of change dm/dt as given by the ratio $m_g/\tau$, where $\tau$ is the time for light to travel a Compton size $\hbar/mc$ one has:

$$\frac{dm}{dt} \approx \frac{\hbar}{c^2 t} \frac{mc^2}{\hbar} = \frac{m}{t} \qquad (10)$$

so that we get by integration

$$m = const \cdot t \qquad (11)$$

and therefore we obtain again the Mass Boom effect. Since $M_u$ and m are proportional to time, the number of particles of cosmological significance in the Universe is constant. The time dependence corresponds to the mass. The above presentation has been submitted to Physics Essays[6].

3.- THE ACTION PRINCIPLE

Einstein's field equations can be derived from an action integral following the Least Action Principle. In standard general relativity one has for the action integral[7]:

$$A = I_G + I_M$$
$$A = -c^3/(16\pi G) \int R\,(g)^{1/2}\,d^4x + I_M \qquad (12)$$

where $I_M$ is the matter action and $I_G$ the gravitational term. Then one obtains the field equations

$$G^{\mu\nu} = 8\pi(G/c^4) \cdot T^{\mu\nu} \qquad (13)$$

We assume a space-time metric and use the Robertson-Walker model that satisfies the Weyl postulate and the cosmological principle, i.e.

$$ds^2 = c(t)^2 \, dt^2 - R^2(t) \{ dr^2 / (1-kr^2) + r^2 ( d\theta^2 + \sin^2\theta \, d\phi^2 ) \} \qquad (14)$$

Einstein's equations (13) follow from the Action (12) provided that the variation of the coefficient in the integral in equation (12) be zero. Then

$$\tilde{c}^3/(16\pi G) = \text{constant} \qquad (15)$$

We see that the assumption of a time varying G must include a time varying c to preserve the form of the field equations.

The equation (15) strongly suggests a specific link between mass and time. This is

$$\tilde{c}^3/G \approx 4.04 \times 10^{38} \text{ grams/sec} = \text{constant} \qquad (16)$$

which is of the order of the ratio of the mass of the observable Universe to its age.

On the other hand, the action for a free material point is

$$A = -mc \int ds \qquad (17)$$

To preserve standard mechanics we make the momentum mc constant, independent from the cosmological time, then

$$mc = \text{constant} \qquad (18)$$

With the constancies expressed in (15) and (18), general relativity is preserved and of course the Newtonian mechanics too. Within these limits time variations of some of the fundamental constants, G, c and masses, are allowed at the same time preserving the laws of physics as we know them today. From (18) and the Mass Boom effect, the speed of light *decreases* linearly with time $c \propto 1/t$. It is evident that, with such a law for the speed of light, the size of the Universe ( of the order of ct ) is constant and therefore there is no "absolute" expansion.

4.-REINTERPRETATION OF THE RED SHIFT: TIME VARIATION OF PLANCK´S "CONSTANT".

The ratio of frequencies observed at the laboratory system, photons from distant galaxies as $\nu = c/\lambda$ and local atomic clocks as $\nu_o \propto mc^2/\hbar$, with mc constant and $\lambda$ also constant (no expansion), gives a red shift proportional to $\hbar$. With no expansion the red shift implies a *decreasing* Planck´s "constant". Now, Planck´s units are defined as a combination of G, c and $\hbar$:

$$\text{Planck´s mass} \quad (\hbar c/G)^{1/2} = 2 \times 10^{-5} \text{ grams}$$

$$\text{Planck´s time} \quad (G\hbar/c^5)^{1/2} = 5.4 \times 10^{-44} \text{ sec}$$

$$\text{Planck´s length} \quad (G\hbar/c^3)^{1/2} = 1.6 \times 10^{-33} \text{ cm} \quad (19)$$

It is evident that if we choose a system of units such that $G = c^3$, as required by the Mass Boom effect on the whole Universe, and such that $\hbar = c^2$, we get Planck´s units of mass and time as the "natural" units of mass and time. This is very appealing because the ratio of the mass and age of the Universe to the corresponding Planck´s units is the same factor of about $10^{61}$. On the other hand the constant size of the Universe, the model we present here, has a value of the order of $ct = 10^{28}$ cm, which is Planck´s length at the first "tic" of time (at Planck´s time).

*We see now that the Boom of an initial fluctuation of time and mass of the Planck´s units, by the factor $10^{61}$, brings the fluctuation up to the state of the Universe as we observe it today in time and mass. On the other hand this initial fluctuation had a size of the order of Planck´s length at that time, which is the constant size of the Universe up to today (about $10^{28}$ cm).*

Then, this factor of $10^{61}$ is representative of the evolution of the initial fluctuation, as characterized by the Planck´s units, followed then by the Mass Boom to bring the Universe to the present conditions. The magic number of the Universe is then $10^{61}$, as representative of its evolution from the initial fluctuation up to now. The cosmology to be studied now in this model is one that keeps $G = c^3$, $\hbar = c^2$ and $ct = 1$.

5.- COSMOLOGICAL EQUATIONS

The Einstein cosmological equations derived from his general theory of relativity are[7]

$$\left(\frac{\dot{a}}{a}\right)^2 + \frac{2\ddot{a}}{a} + 8\pi G \frac{p}{c^2} + \frac{kc^2}{a^2} = \Lambda c^2$$

$$\left(\frac{\dot{a}}{a}\right)^2 - \frac{8\pi}{3} G\rho + \frac{kc^2}{a^2} = \frac{\Lambda c^2}{3} \qquad (20)$$

The solution for $\hbar = c^2$, as presented in the previous section, implies a redshift given by an apparent value of $a(t) \propto t^2$. In the units we have selected, consistent with this interpretation of the redshift as a decrease in $\hbar$, the curvature term in (20) decreases as $t^{-4}$ so that it is negligible, and we are observing essentially a flat Universe. With the present reasonable approximation of zero pressure (neglecting random speeds of galaxies), and substituting $a(t) \propto t^2$ in (20) we finally get the cosmological equations:

$$\left(\frac{2}{t}\right)^2 + \frac{4}{t^2} = \Lambda c^2$$

$$\left(\frac{2}{t}\right)^2 - \frac{8\pi}{3}G\rho = \frac{\Lambda c^2}{3} \quad (21)$$

We convert now these equations to the standard definitions:

$$\Omega_m = \frac{8\pi}{3}\frac{G\rho a^2}{\dot{a}^2} = \frac{8\pi}{3}G\rho\frac{t^2}{4}$$

$$\Omega_\Lambda = \frac{\Lambda c^2}{3}\frac{a^2}{\dot{a}^2} = \frac{\Lambda c^2}{3}\frac{t^2}{4} \quad (22)$$

and therefore we get

$$\Omega_m = 1/3$$

$$\Omega_\Lambda = 2/3 \quad (23)$$

These numbers are very close to the current values observed at present. The accelerated expansion of the Universe[8] is then an apparent effect due to the quadratic relation $a(t) \propto t^2$ as seen from the laboratory system.

6.- ENTROPY OF THE UNIVERSE: LINEAR WITH TIME

We have proved elswhere[4] that the entropy of the Universe varies linearly with cosmological time, based upon a new approach. However, using the well known Bekenstein[9] and Hawking[10] relations for entropy, as well as the classical definition, the result is the same: there is no escape, the entropy varies linearly with time and for the Universe the high entropy of today is due to the fact that the Universe is very old. ***There is no entropy problem in our model.***

Boltzmann constant k varies in our theory as c. To see this we have the photon relation typical for blackbody radiation

$$kT \propto \hbar c/\lambda \qquad (24)$$

Taking the laboratory system $\hbar$ is constant and from the empirical law $T\lambda$ = constant we get k varying as c, inversely proportional to cosmological time. The apparent time variation of T is $T \propto 1/\lambda \propto 1/a(t)$, which is the classical relation used in many cosmological model. Hence the Bekenstein definition of entropy:

$$S/k \propto \text{Energy x size} / \hbar c \propto Mc^2 \times (ct)/\hbar c \propto t^2$$

$$\text{gives } S \propto t \qquad (25)$$

For the Hawking black hole entropy: $S/k \propto 1/\hbar c \cdot (GM^2) \propto 1/\hbar = t^2$ i.e. the same result. For the standard $S = \text{Energy}/T \propto Mc^2 \times a(t) \propto t$ we also get the same result.

## 7.- THE MAGIC NUMBERS

The only magic number we found here is $10^{61}$ that brings the first Planck fluctuation to the present state of the Universe. The Dirac magic number $10^{40}$, as the ratio of the size of the Universe to the size of fundamental constants, and the ratio of electric to gravitational forces, is a function of time in our approach here. Therefore the similarity of these two values is a coincidence in our interpretation. Weinberg´s relation[7], that can be derived by equating the gravitational cross section (6) of a particle of mass m to the square of its Compton wavelength, is

$$\hbar^2/(Gct) \approx m^3 \qquad (26)$$

The time dependence implied here for a typical mass m of a particle is $m^3 \propto 1/t$ which has no meaning in our approach. But at the Lab system we have $\hbar =$ constant and then we get from (20) that m is proportional to time t, again the Mass Boom is also present here.

8.- PREDICTIONS

Using the expression of the fine structure constant found with no c in it elsewhere[6] we get

$$\alpha \approx e^2/\hbar = (e/c)^2 \qquad (27)$$

There have been no cosmologically significant time variations in $\alpha$, by that meaning variations of the order of the variation of the cosmological age considered. Then one must have e/c = constant, and therefore the electronic charge e varies as c, inversely proportional to t. However, in electromagnetic units (e/c) is a true constant, so that the Zeemann displacement is a constant in this theory, contrary to the statement we made elsewhere[1].

The apparent Hubble "constant" in this theory is H = 2/t , due to the cosmological scale factor varying as $t^2$. Hence the Hubble age in this theory is twice as much as the standard one. It is suggested that the age of the Universe may be as much as twice what we have been thinking up to now.

Finally the Pioneer[3] 10/11 anomalous acceleration observed can be explained here by the ratio of the laboratory system reference frequency ($\hbar$ = constant) $\nu_l$

$$\nu_l \propto mc^2/\hbar \propto c = 1/t \qquad (28)$$

and the frequency $\nu_p$ of the photon observed ( $\hbar = c^2$ )

$$\nu_p \propto mc^2/\hbar \propto m \propto t \qquad (29)$$

Hence we have $\nu_p/\nu_l \propto t^2$. This is a **BLUE SHIFT**, as observed, and of the order of $Hc \approx 7 \times 10^{-8}$ cm/sec$^2$ to be compared with the observed value of about $8 \times 10^{-8}$ cm/sec$^2$. Our theory explains the anomalous acceleration observed in the Pioneer 10/11 probes.

9.- CONCLUSIONS

The Mass Boom proposed, linear increase of all masses with time, implies here a linear decrease of the speed of light. The resultant cosmological model, static, almost flat, closed and finite, has cosmological parameters in accordance with current observations. The main problems of the standard model are solved: entropy, lambda constant, horizon etc. In fact many of these problems are one and the same thing. Solving one you solve all of them. This is the case here.

Finally, the time reversibility of all the equations of physics poses a deep theoretical problem: nature has irreversible process, and this irreversibility is not now explicit in the standard basic equations of physics (Newton´s mechanics, quantum mechanics, general relativity, etc). With our approach the Mass Boom ensures that irreversibility is present everywhere: in fact we have proved that it corresponds to an Entropy Boom linear with time. On our theory irreversibility comes from the basic process of gravity quanta emission, responsible for the Mass Boom, the time Boom, the Entropy Boom.

# 10.-REFERENCES